\documentclass[a4paper]{article}

\usepackage{INTERSPEECH2021}

\usepackage[dvipdfmx]{graphicx}

\usepackage{bm}
\usepackage{epsfig} 
\usepackage{mathptmx} 
\usepackage{times} 
\usepackage{amsmath} 
\usepackage{amssymb}  
\usepackage{algorithm}
\usepackage{algorithmic}
\usepackage{url}
\usepackage{cite}
\usepackage{diagbox}

\title{StarGAN-based Emotional Voice Conversion for Japanese Phrases}
\name{Asuka Moritani$^1$, Ryo Ozaki$^2$, Shoki Sakamoto$^2$, Hirokazu Kameoka$^3$, Tadahiro Taniguchi$^1$}
\address{
  $^1$College of Information Science and Engineering, Ritsumeikan University, Japan\\
  $^2$Graduate School of Information Science and Engineering, Ritsumeikan University, Japan\\
  $^3$NTT Communication Science Laboratories, NTT Corporation, Japan}
\email{\{moritani.asuka, taniguchi, ryo.ozaki, sakamoto.shoki\}@em.ci.ritsumei.ac.jp, hirokazu.kameoka.uh@hco.ntt.co.jp}

\begin{document}
\maketitle

\begin{abstract}
This paper shows that StarGAN-VC, a spectral envelope transformation method for non-parallel many-to-many voice conversion (VC), is capable of emotional VC (EVC). Although StarGAN-VC has been shown to enable speaker identity conversion, its capability for EVC for Japanese phrases has not been clarified. In this paper, we describe the direct application of StarGAN-VC to an EVC task with minimal fundamental frequency and aperiodicity processing. Through subjective evaluation experiments, we evaluated the performance of our StarGAN-EVC system in terms of its ability to achieve EVC for Japanese phrases. The subjective evaluation is  conducted in terms of subjective classification and mean opinion score of neutrality and similarity.
In addition, the interdependence between the source and target emotional domains was investigated from the perspective of the quality of EVC. 
\end{abstract}

\section{Introduction}\label{sec:1}
Emotional expression in our daily conversations is essential for conveying our intentions, emotions, and social attitudes. Therefore, generating speech signals with designated emotional attributes is an important objective in the speech synthesis field.
Producing emotional utterances artificially  and converting a source utterance with a particular emotional attribute to one that is different is an important task. This task is termed emotional voice conversion (EVC).
In this study, we explore the StarGAN-VC's  ability of EVC for Japanese phrases.

EVC has long been studied~\cite{mori2006emotional,tao2006prosody,inanoglu2007system,aihara2012gmm,ming2015fundamental,Ming16-DBL,Luo19-EVC}. Recently, Ming et al. developed a method using a deep bidirectional LSTM model~\cite{Ming16-DBL}, and Luo et al. proposed a method using dual supervised adversarial networks~\cite{Luo19-EVC}. Prosody is widely considered an important feature for expressing emotion. Therefore, most EVC studies and techniques focus mainly on the conversion of fundamental frequency ($F_0$) contours. 
In this paper, instead of focusing on $F_0$ and aperiodicity transformation, we investigate whether EVC can be achieved by spectral envelope transformation alone.

In addition, many existing EVC methods require a parallel corpus of utterances with different emotional attributes. However, preparing such a corpus can be costly and not always feasible. In this respect, a method that does not require a parallel corpus and can be trained even with a non-parallel corpus would be very useful.

For speaker identity conversion, a method based on a variant of generative adversarial networks (GANs) called StarGAN \cite{Choi2018stargan} was proposed~\cite{Kameoka18-SNM}. This method, called StarGAN-VC, has been shown to be capable of non-parallel many-to-many VC (i.e., learning mappings between multiple speakers' voices using a non-parallel corpus). StarGAN-VC was developed as an extension of CycleGAN-VC~\cite{Kaneko2018cyclegan}, which was designed to handle non-parallel one-to-one VC. 
 In this paper, we study StarGAN-VC applied to the EVC task and call it StarGAN-based EVC (StarGAN-EVC).
In StarGAN-EVC, we focus on spectral envelope transformation only. For $F_0$ and aperiodicity, we use simple methods, i.e., logarithm  Gaussian normalization for $F_0$~\cite{Liu2007} and identity mapping for the aperiodicity. Namely, the aim of this study is to investigate how well StarGAN-VC can achieve EVC when it is applied only to spectral envelope transformation.
%
Note that a similar study on English speech had already been done~\cite{gao2019nonparallel}. However, to the best of our knowledge, this is the first study on StarGAN-VC-based EVC evaluated on Japanese speech.
Thus, the main contribution of this paper is the subjective evaluation experiment conducted to show that StarGAN-VC is capable of EVC for Japanese phrases. 

The remainder of this paper is organized as follows. Section~\ref{sec:2} describes the original formulation of StarGAN-VC, and Section~\ref{sec:3} describes the detail of StarGAN-EVC. Section~\ref{sec:4} describes the experiment. Finally, Section~\ref{sec:5} concludes this paper with a discussion. 

\section{Preliminaries: StarGAN-VC}\label{sec:2}
\begin{figure}[tb!p]
  \begin{center}
    \includegraphics[width=0.9\linewidth]{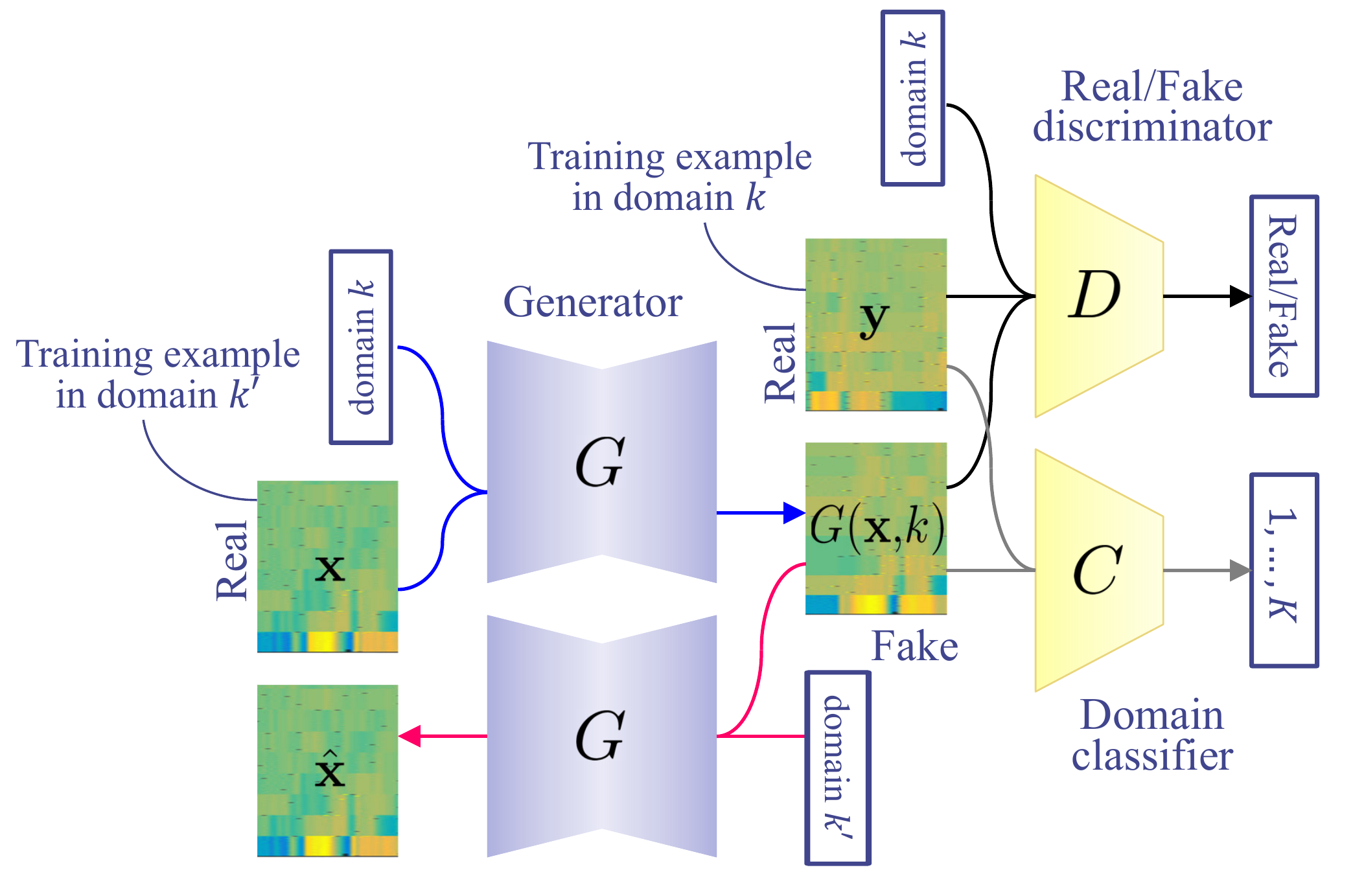}    \vspace{-5mm}
    \caption{Concept figure of StarGAN-VC~\cite{Kameoka18-SNM}}
    \label{fig:StarGAN-VC_concept}
  \end{center}
  \begin{center}
    \includegraphics[width=0.75\linewidth]{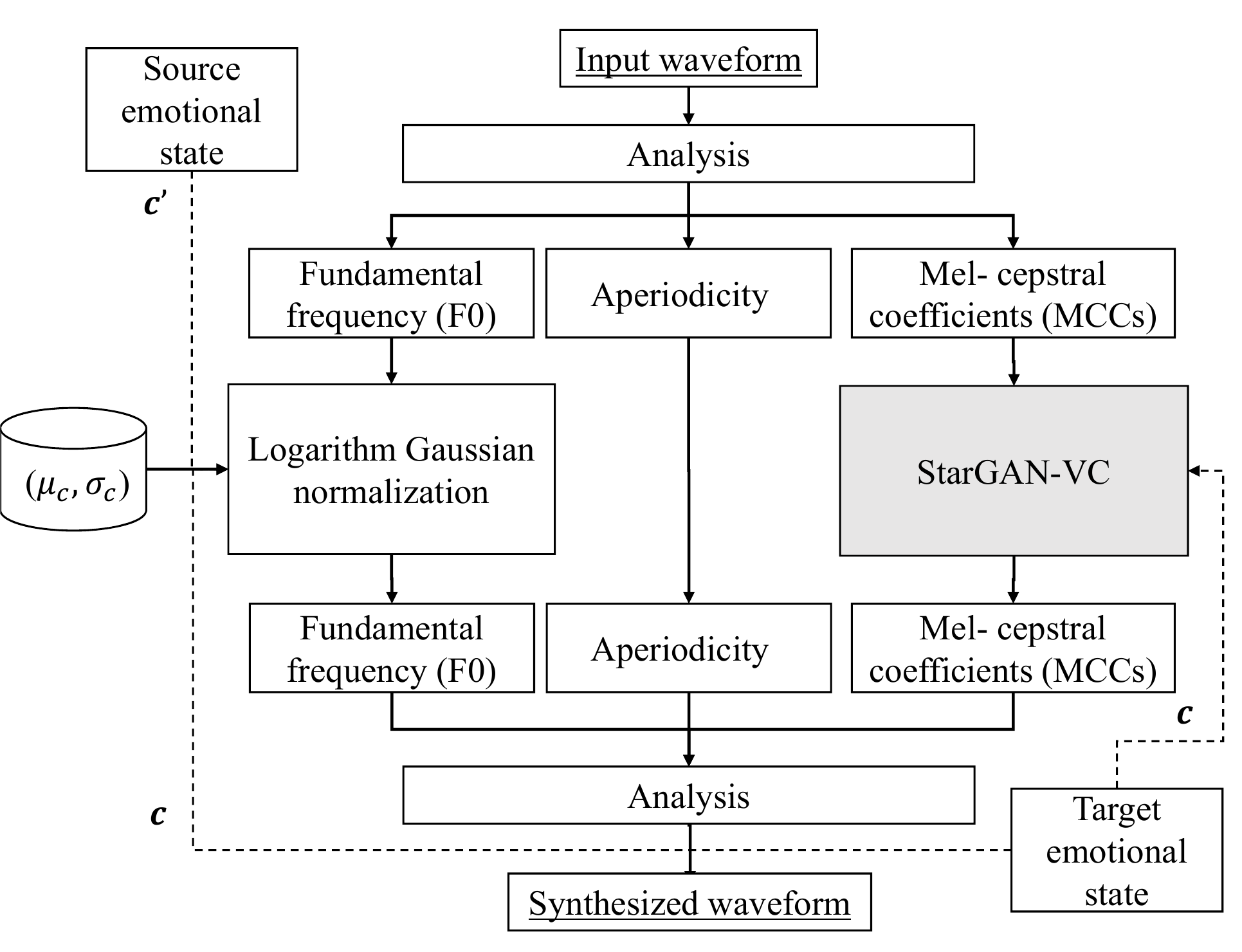}
    \vspace{-3mm}
    \caption{Overview of StarGAN-Emotional-VC}
    \label{fig:StarGAN-Emotional-VC}
  \end{center}\vspace{-5mm}
\end{figure}

StarGAN-VC is a non-parallel and many-to-many voice conversion method based on a variant of GANs called StarGAN. 
The overview of the method is shown in Figure~\ref{fig:StarGAN-VC_concept}.
StarGAN-VC consists of three networks, a generator, real/fake discriminator, and domain classifier. 
The generator takes a speech feature sequence in the source domain and the index $c$ of the target domain as the inputs and generates a converted version of the feature sequence. At training time, the feature sequence generated by the generator is evaluated by the real/fake discriminator and domain classifier.
The generator and discriminator are trained in an adversarial manner. 
Namely, while the discriminator is trained to correctly distinguish the feature sequences generated by the generator from those of real speech, the generator is trained to deceive the discriminator by making the generated feature sequences as indistinguishable as possible from the feature sequences of real speech.
In addition, the generator and domain classifier are trained to cooperatively maximize the classification performance of the classifier. Namely, the generator is trained to generate feature sequences so that they are correctly classified by the classifier as belonging to the target domain.

Let us use $\hat{\bm{y}}=G(\bm{x}, c)$ to express the output of the generator $G$, where $\bm{x}\in\mathbb{R}^{Q \times N}$ denotes an input feature sequence of dimension $Q$ and length $N$, and $c$ denotes the target domain index (the index of the domain in which $G$ attempts to convert $\bm{x}$).
Here, $c$ is assumed to be represented as a one-hot vector.
The discriminator $D$ outputs the probability $D(\bm{y},c)\in [0,1]$ of an input feature sequence $\bm{y}$ being ``real'', and the domain classifier $C$ outputs the probability $p_{C}(c|\bm{y})\in [0,1]$ of an input feature sequence $\bm{y}$ belonging to domain $c$. 
By using these expressions, the training losses are defined as follows.



\textbf{Adversarial loss}: The losses for the generator $G$ and discriminator $D$ can be defined as follows:  
\begin{align}
  \mathcal{L}^{D}_\mathrm{adv}(D)
  &= -\mathbb{E}_{c\sim p(c),\bm{y}\sim p(\bm{y}|c)}[\log D(\bm{y},c)]\nonumber\\
  &\ \ \ -\mathbb{E}_{\bm{x}\sim p(\bm{x}),c\sim p(c)}[\log (1-D(G(\bm{x},c),c))],\nonumber \\
  \mathcal{L}^{G}_\mathrm{adv}(G)
  &=-\mathbb{E}_{\bm{x}\sim p(\bm{x}),c\sim p(c)}[\log D(G(\bm{x},c),c)],\nonumber 
\end{align}
where $\bm{y}\sim p(\bm{y}|c)$ denotes the feature sequence of a ``real'' training utterance belonging to domain $c$, and $\bm{x}\sim p(\bm{x})$ denotes the feature sequence of a training utterance belonging to an arbitrary domain.
The value of $\mathcal{L}^{D}_\mathrm{adv}(D)$ becomes small when $D$ correctly distinguishes ``fake'' samples $G(\bm{x},c)$ from ``real'' samples $\bm{y}$. 
Conversely, the value of $\mathcal{L}^{G}_\mathrm{adv}(G)$ becomes small when the generated samples $G(\bm{x},c)$ are misclassified as ``real'' by $D$. 
Therefore, during training, we would like to minimize 
$\mathcal{L}^{D}_\mathrm{adv}(D)$ with respect to $D$ and 
$\mathcal{L}^{G}_\mathrm{adv}(G)$ with respect to $G$.


\textbf{Domain classification loss}: The above losses are only responsible for encouraging the generated samples $G(\bm{x},c)$ look ``real''. To further encourage the generated samples $G(\bm{x},c)$ to belong to domain $c$, the following training losses for the domain classifier $C$ and $G$ can be used. 
\begin{align}
  &\mathcal{L}^{C}_\mathrm{cls}(C)= -\mathbb{E}_{c\sim p(c),\bm{y}\sim p(\bm{y}|c)}[\log p_{C}(c|\bm{y})],\nonumber \\
  &\mathcal{L}^{G}_\mathrm{cls}(G)=-\mathbb{E}_{\bm{x}\sim p(\bm{x}),c\sim p(c)}[\log p_{C}(c|G(\bm{x},c))].\nonumber 
\end{align}
The values of $\mathcal{L}^{C}_\mathrm{cls}(C)$ and $\mathcal{L}^{G}_\mathrm{cls}(G)$ become small when $C$ correctly classifies $\bm{y}\sim p(\bm{y}|c)$ and $G(\bm{x},c)$ as belonging to domain $c$. Therefore, during training, we would like to minimize $\mathcal{L}^{C}_\mathrm{cls}(C)$ with respect to $C$ and  $\mathcal{L}^{G}_\mathrm{cls}(G)$ with respect to $G$. 


\textbf{Cycle-consistency loss}: Training $G$ using only the adversarial and domain classification losses does not ensure that the linguistic content of an input of $G$ is preserved. Like CycleGAN~\cite{Kaneko2018cyclegan}, the following cycle-consistency loss can be used to encourage $G$ to learn linguistic-preserving conversions
\begin{align}
  \mathcal{L}_\mathrm{cyc}(G) 
  &=\mathbb{E}_{c^{'}\sim p(c),\bm{x}\sim p(\bm{x}|c'),c\sim p(c)}\left[||G(G(\bm{x},c),c')-\bm{x}||_\rho \right]. \label{eq:cycle}
\end{align}
Here, $\bm{x}\sim p(\bm{x}|c')$ denotes the feature sequence of a training utterance belonging to domain $c'$. 
$|\cdot|_\rho$ denotes $L_\rho$-norm where $\rho$ is a positive number, typically set to $\rho = 1$ or $2$.

\textbf{Identity mapping loss}: To let $G$ not make any changes to an input when $c=c'$, the identity mapping loss is introduced as follows: 
\begin{align}
  \mathcal{L}_\mathrm{id}(G)=\mathbb{E}_{c'\sim p(c),\bm{x}\sim p(\bm{x}|c')}[||G(\bm{x},c')-\bm{x}||_\rho ].\nonumber 
\end{align}


\textbf{Overall loss function}: By using the above losses, the overall loss functions for the generator, discriminator, and domain classifier can be defined as follows: \begin{align}
  &\mathcal{I}_{G}(G)=\mathcal{L}^{G}_\mathrm{adv}(G)+\lambda_\mathrm{cls}\mathcal{L}^{G}_\mathrm{cls}(G)
  +\lambda_\mathrm{cyc}\mathcal{L}_\mathrm{cyc}(G)+\lambda_\mathrm{id}\mathcal{L}_\mathrm{id}(G),\nonumber \\
  &\mathcal{I}_{D}(D)=\mathcal{L}^{D}_\mathrm{adv}(D), \ \ \ \mathcal{I}_{C}(C)=\mathcal{L}^{C}_\mathrm{cls}(C),\nonumber 
\end{align}
where the hyperparameters $\lambda_\mathrm{cls}\geq 0$ and $\lambda_\mathrm{cyc}\geq 0$, $\lambda_\mathrm{id}\geq 0$ represent the importance of the domain classification, cycle-consistency, and identity mapping losses, respectively.
These hyperparameters can be set manually or optimized by hyperparameter search.

For more details, please refer to the original paper~\cite{Kameoka18-SNM}.

\section{StarGAN-EVC}\label{sec:3}

In StarGAN-EVC, an emotional state is used as the conditioning input $c$.
Figure~\ref{fig:StarGAN-Emotional-VC} illustrates the overall StarGAN-EVC system. 
First, an input waveform is transformed into the sequences of the $F_0$s, mel-cepstral coefficients (MCCs), and aperiodicities, each extracted within a short-term frame.
The sequence of the MCC vectors is then fed into the StarGAN-VC module. In StarGAN-EVC, the domain $c$ corresponds to an emotional state (e.g., joyful, sad, or surprised).
At training time, training utterances labeled with emotional states are used. 
At test time, the MCC sequence of input speech is converted via the generator using the target emotional state label $c$.

The $F_0$ contour of input speech is converted using logarithm  Gaussian normalization~\cite{Liu2007}.
Logarithm Gaussian normalization is a method that
simply adjusts the entire $F_0$ contour using a linear transformation
in the logarithmic domain
\begin{align}
\log(f_{\rm converted})&=\frac{\sigma_{c}(\log(f_{\rm input}) - \mu_{c'})}{\sigma_{c'}}+\mu_{c} \nonumber 
\end{align}
where $\mu_{c'}$ and $\sigma_{c'}$ are the mean and standard deviation of the logarithmic $F_0$s of the source emotional state, and $\mu_{c}$ and $\sigma_{c}$ are those of the target emotional state.
The statistical parameters $\mu_{c}$, $\sigma_{c}$, $\mu_{c'}$, and $\sigma_{c'}$ are calculated using the training utterances. 

The aperiodicity sequence extracted from input speech is directly used without modification.

\section{Experiment}\label{sec:4}
\subsection{Experimental setup}
We used the Keio University Japanese Emotional Speech Database  (Keio-ESD)\footnote{Keio University Japanese Emotional Speech Database: \url{http://research.nii.ac.jp/src/en/Keio-ESD.html}} developed by Moriyama~\cite{Moriyama09}, and evaluated the StarGAN-Emotional-VC by subjective evaluation.
The acoustic signals recorded in Keio-ESD were directly recorded to the hard disk drive by a microphone connected to the PC in a soundproof room. 
A 32-year old male speaker who had acting experience spoke 20 short Japanese phrases expressing 47 emotional states. The spoken phrases recorded as Waveforms were digitized by 16-kHz sampling and 16-bit quantization. The data were stored in the WAV format.

For analyzing the input waveforms and synthesizing the waveforms of the converted acoustic features, we used WORLD~\cite{Morise16-WAV} (D4C edition~\cite{Morise16-DAB}), which is a high-quality speech analysis and synthesis system akin to a vocoder.   

Logarithmic fundamental frequency (log $F_0$), spectral envelope, and aperiodicity were extracted every $5$ ms using WORLD~\cite{Morise16-WAV} (D4C edition~\cite{Morise16-DAB}.
For training StarGAN-VC, $36$-dimensional MCCs calculated from the spectral envelope were used.

In this experiment, we adopted four representative emotional states.
They were neutrality labeled (neutral) and three other typical emotional states, i.e., joy (joyful), anger (angry), and sorrow (sad), which are distant from each other in the circumplex model of the affect grid proposed by Russell~\cite{Russell80-ACM}.

We employed the same network architecture as that of the original paper~\cite{Kameoka18-SNM}. Namely, gated convolutional neural networks were used for the generator, the discriminator and the domain classifier. 
For training StarGAN-VC, 17 of the 20 selected phrases were used as a non-parallel corpus for the training dataset. The remaining three phrases, ``amamizuwa,'' ``midori,'' and ``nami''\footnote{They mean ``rainwater is,'' ``green,'' and ``wave'' in English, respectively}, were used as test data.
In the StarGAN-VC training phase, the batch size and the number of epochs were set to $2$ and $2000$, respectively. We used Adam optimizer with $(\alpha. \beta_1 )=(1.0\times 10^{-3}, 0.9)$, $(5.0\times 10^{-5}, 0.5)$, and $(1.0\times 10^{-3}, 0.5)$ for the generator, domain classifier, and adversarial real/fake discriminator, respectively. Other parameters for Adam were set to the default values~\cite{Kingma2014adam}.

We performed emotional voice conversions using StarGAN-Emotional-VC.
Test data with the ``neutral'' label were converted into ``joyful,'' ``angry,'' and ``sad'' speech.
Those labeled ``joyful'' were converted into ``neutral,'' ``angry,'' and ``sad'' speech.
Those labeled ``angry'' were converted into ``neutral,'' ``joyful,'' and ``sad'' speech.
Those labeled ``sad'' were converted into ``neutral,'' ``joyful,'' and ``angry'' speech.
In total, we converted 36 synthesized speech signals.

The 48 data units included 12 original test data units, i.e., 3  phrases with 4 emotional expressions and 36 synthesized data units, i.e., 3 phrases with 4 emotional expressions converted from 4 source input data corresponding to 4 emotional states.

\subsection{Subjective evaluation }
\subsubsection{Subjective classification}
Ten participants were asked to label a converted voice as ``neutral,'' `joyful,'' ``angry,'' or ``sad'' \footnote{The participants were 21--24 year old undergraduate and graduate students and native Japanese speakers.}. The 48 audio data units were played randomly to each participant.

\begin{table}[bt]
  \centering
  \caption{Subjective classification results for the original input data}
  \label{tbl:original_eval}
  \begin{tabular}{|c||c|c|c|c|} \hline
 Category &\multicolumn{4}{c|}{Subjective classification}\\
     & Neutral & Joyful & Angry & Sad \\ \hline \hline
    Neutral & \underline{\bf{100.0\%}} & 0.0\% & 0.0\% & 0.0\% \\ \hline
    Joyful & \underline{26.7\%} & \underline{\bf{56.7\%}} & 16.7\% & 0.0\% \\ \hline
    Angry & \underline{10.0\%} & 0.0\% & \underline{\bf{90.0\%}} & 0.0\% \\ \hline
    Sad & 0.0\% & 0.0\% & 0.0\% & \underline{\bf{100.0\%}} \\ \hline
  \end{tabular}
  \centering
  \caption{Subjective classification results for the converted data}
  \label{tbl:response_rate}
  \begin{tabular}{|c|c||c|c|c|c|} \hline
  
  \multicolumn{2}{|c||}{Domain}&\multicolumn{4}{c|}{Subjective classification}\\
Source	&	Target	&	Neutral	&	Joyful	&	Angry	&	Sad	\\\hline\hline
	&	Joyful	&	\underline{\bf{46.7}}\%	&	\underline{30.0}\%	&	13.3\%	&	10.0\%	\\\cline{2-6}
Neutral	&	Angry	&	\underline{\bf{43.3}}\%	&	10.0\%	&	\underline{\bf{43.3}}\%	&	3.3\%	\\\cline{2-6}
	&	Sad	&	13.3\%	&	\underline{23.3}\%	&	10.0\%	&	\underline{\bf{53.3}}\%	\\\hline\hline
	&	Neutral	&	\underline{\bf{63.3}}\%	&	6.7\%	&	\underline{20.0}\%	&	10.0\%	\\\cline{2-6}
Joyful	&	Angry	&	\underline{26.7}\%	&	16.7\%	&	\underline{\bf{50.0}}\%	&	6.7\%	\\\cline{2-6}
	&	Sad	&	3.3\%	&	\underline{20.0}\%	&	16.7\%	&	\underline{\bf{60.0}}\%	\\\hline\hline
	&	Neutral	&	\underline{\bf{83.3}}\%	&	\underline{6.7}\%	&	3.3\%	&	\underline{6.7}\%	\\\cline{2-6}
Angry	&	Joyful	&	13.3\%	&	\underline{\bf{50.0}}\%	&	\underline{36.7}\%	&	0.0\%	\\\cline{2-6}
	&	Sad	&	20.0\%	&	\underline{26.7}\%	&	\underline{\bf{36.7}}\%	&	16.7\%	\\\hline\hline
	&	Neutral	&	\underline{20.0}\%	&	0.0\%	&	6.7\%	&	\underline{\bf{73.3}}\%	\\\cline{2-6}
Sad	&	Joyful	&	\underline{16.7}\%	&	\underline{16.7}\%	&	13.3\%	&	\underline{\bf{53.3}}\%	\\\cline{2-6}
	&	Angry	&	\underline{16.7}\%	&	10.0\%	&	13.3\%	&	\underline{\bf{60.0}}\%	\\\hline
  \end{tabular}
  \centering
  \caption{Averaged response rate of emotional attributes of converted speech regardless of attribute of input data}
  \label{tbl:average}
  \begin{tabular}{|c||c|c|c|c|} \hline
    \multicolumn{1}{|c||}{Domain}&\multicolumn{4}{c|}{Subjective classification}\\
	Target	&	Neutral	&	Joyful	&	Angry	&	Sad	\\\hline\hline
    Neutral  & \underline{\bf{55.6\%}} & 4.4\% & 10.0\% & \underline{30.0\%} \\ \hline
    Joyful
 & \underline{25.6\%} & \underline{\bf{32.2\%}} & 21.1\% & 21.1\% \\ \hline
    Angry & \underline{28.9\%} & 12.2\% & \underline{\bf{35.6\%}} & 23.3\% \\ \hline
    Sad & 12.2\% & \underline{23.3\%} & 21.1\% & \underline{\bf{43.3\%}} \\ \hline
  \end{tabular}
\end{table}

Table~\ref{tbl:original_eval} shows the subjective classification results for the original input data, i.e., test data.
The highest values are bold and underlined, and the second-highest values are underlined, except for $0.0\%$, in the following tables.  
Each rate means the ratio of the participants who did or did not identify the emotional category when they listened to the phrase included in the set of test data for each emotional state label.

Table~\ref{tbl:original_eval} shows that the participants almost all correctly identified the emotion categories by listening to the data, although ``joyful'' included over $40\%$ misrecognitions.
This means that the four emotional states were evoked correctly in the test data and the participants were capable of evaluating the synthesized data.

Table~\ref{tbl:response_rate} shows the subjective classification results for the converted data with respect to the source and target attribute domains. The results averaged over every source attribute domain are shown in Table~\ref{tbl:average}.

This demonstrates that, generally, StarGAN-Emotional-VC could, to a certain extent, convert neutral utterances into utterances from each emotional category.
With every result, the classification rate of the source attribute domain, i.e., emotion category, decreased, and that of the target attribute domain increased.
In particular, in the conversions from ``neutral'' to ``angry'' and ``sad,'' from ``joyful'' to another emotional state, and from ``angry'' to ``neutral'' and ``joyful,'' the target attribute domain successfully obtained the highest correctness score.

In contrast, every synthesized utterance converted from ``sad'' remained in the ``sad'' emotional state, i.e., the subjective classification rate of ``sad'' received the highest score.
We consider that this is because aperiodicity is known to contribute a ``sad'' overtone expression compared to other emotional expressions~\cite{Scherer1986-VAE}. Considering that StarGAN-Emotional-VC converted only MCCs, this result is understandable.   

However, collectively, Table~\ref{tbl:average} shows averaged subjective classification rate of an emotional attribute of converted speech regardless of the attribute of input data. This shows that the emotional voice conversion successfully generated utterances with target emotional expressions.

The result shows that StarGAN-VC could convert an emotional expression to a certain extent even though it does not apply any modern techniques to the conversion of $F_0$ and aperiodicity features.

\subsubsection{Mean opinion and emotional similarity scores}

\begin{figure}[t]
  \begin{center}
    \includegraphics[width=0.7\linewidth]{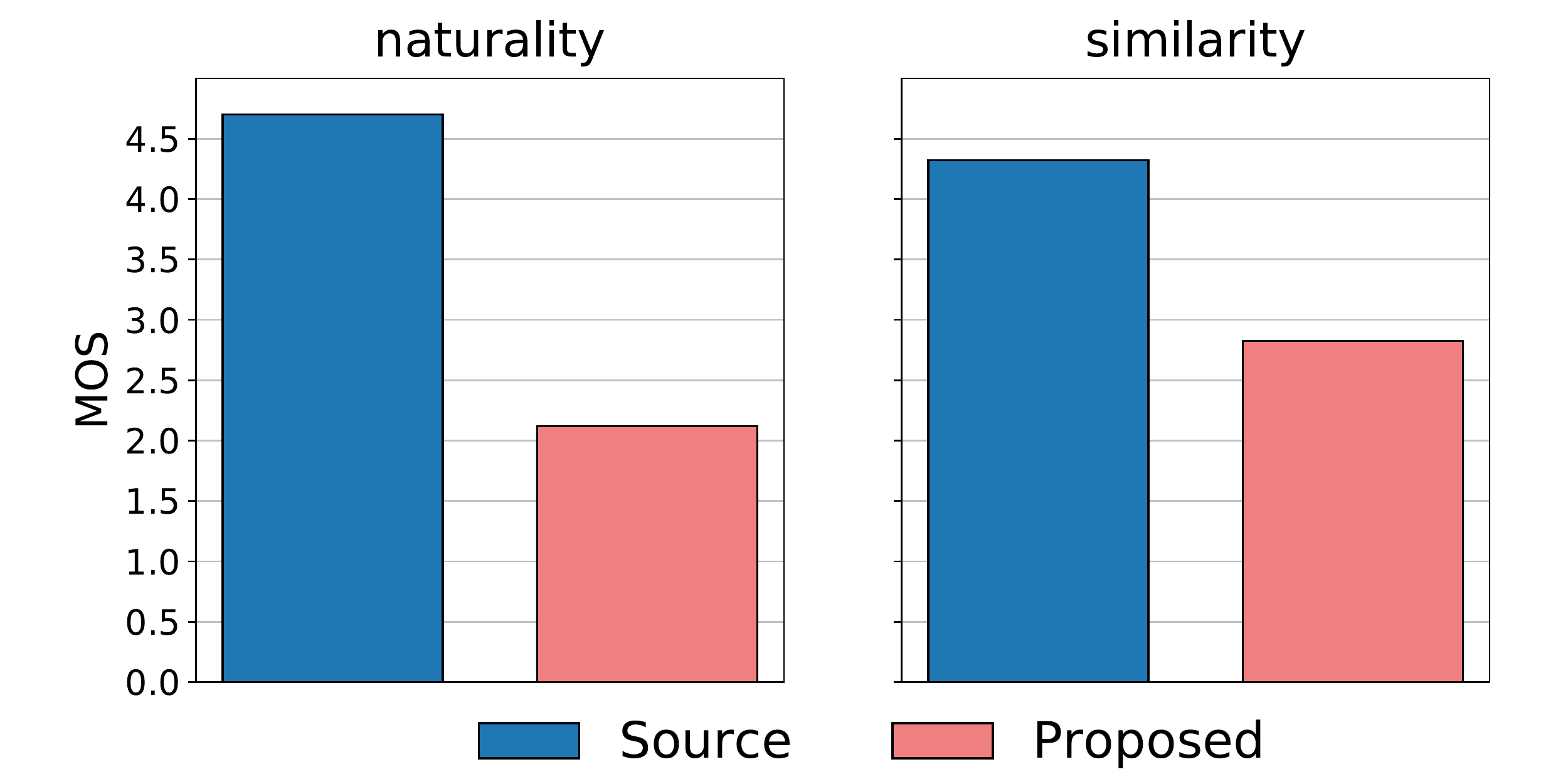}
    \caption{Result of average on all MOS and ESS}
    \label{M_result}
  \end{center}
\end{figure}

\begin{table}[bt]
  \centering
  \caption{Result of MOS (for naturalness)}
  \label{tbl:naturalness}
  \begin{tabular}{|c||c|c|c|c||c|} \hline
    \multicolumn{1}{|c||}{Domain}&\multicolumn{4}{c||}{Domain Source}&\multicolumn{1}{c|}{Original}\\
	Target	&	Neutral	&	Joyful	&	Angry	&	Sad  &	\\\hline\hline
    Neutral &   --  & 2.50 & 2.50 & 1.92 & 4.79 \\ \hline
    Joyful  & 2.13 &  --  & 2.50 & 1.50 & 4.83 \\ \hline
    Angry   & 2.13 & 2.00 &  --  & 1.58 & 4.46 \\ \hline
    Sad     & 1.88 & 2.71 & 1.96 &  --  & 4.79 \\ \hline
  \end{tabular}
  \centering
  \caption{Result of ESS (for similarity)}
  \label{tbl:similarity}
  \begin{tabular}{|c||c|c|c|c||c|} \hline
    \multicolumn{1}{|c||}{Domain}&\multicolumn{4}{c||}{Domain Source}&\multicolumn{1}{c|}{Original}\\
	Target	&	Neutral	&	Joyful	&	Angry	&	Sad  &	\\\hline\hline
    Neutral  & --  & 3.17 & 3.25 & 3.04 & 4.67 \\ \hline
    Joyful   & 2.54 & --  & 2.42 & 1.92 & 3.33 \\ \hline
    Angry    & 2.63 & 3.08 &  --  & 2.04 & 4.54 \\ \hline
    Sad      & 3.58 & 3.50 & 2.92 &  --  & 4.92 \\ \hline
  \end{tabular}
\end{table}
 
Mean opinion score (MOS) and emotional similarity score (ESS) tests were also performed to determine whether the converted speech signals were natural and similar to the target speech signals corresponding to each emotional state, respectively. Eight participants were asked to evaluate the converted speech signals.
Each participant was requested to listen to and evaluate each converted speech signal on a 5-point scale.
The converted speech signal was evaluated by participants based on how natural and similar it was to the target voice. As a result, we obtained MOS for ``naturalness'' and ESS for  ``similarity''.

Figure~\ref{M_result} shows the averaged MOS and ESS scores for the original and converted voices. It shows that MOS for naturalness was lower than ESS for similarity. 
This means the overall quality of converted voice has room for improvement though the emotional conversion was achieved to a certain extent\footnote{It should be noted that the dataset size was small because of the scarcity of available datasets for emotional voice conversion. Increasing the size of the dataset may improve naturalness to a certain extent.}. 

Tables~\ref{tbl:naturalness} and \ref{tbl:similarity} show the MOS for ``naturalness'' and ESS for ``similarity'' for each pair of emotional states. 
The final columns of the tables show MOS or ESS of the original voice, i.e., non-converted voice, as references. 
The value of each cell is the average of 24 scores, i.e., eight participants for each utterance and three utterances for each emotional state. 
The tendencies we found in the tables, in general, share the same characteristics as the classification-based evaluation. However, some interesting features were observed. 

When the source domain was ``sad'', the MOS for each target domain was less than $2$. This means that the system, i.e., GAN-based spectral envelope transformations, could not produce a natural voice. This suggests that ``sad'' has a very different spectral envelope from the viewpoint of GAN-based transformations and typical characteristics of other features, e.g., aperiodicity and speech speed.

Furthermore, it is worth noting that Tables~\ref{tbl:naturalness} and \ref{tbl:similarity} have different tendencies and asymmetricity. Regarding a pair of emotional states, high ESS does not mean high MOS, and the possibility of voice conversion from emotional state A to B does not guarantee that from B to A despite the contribution of cycle-consistency loss (see (\ref{eq:cycle})) in StarGAN-VC. This suggests that transformation of other features, e.g., speech speed, is crucial for bi-directional emotional voice conversion. 


\section{Conclusion}\label{sec:5}
This study conducted emotional voice conversion using StarGAN-VC, i.e., StarGAN-Emotional-VC, and evaluated the capability of emotional voice conversion for Japanese phrases through a subjective listening test.
The result shows that StarGAN-VC can perform emotional voice conversion mainly through spectral envelope transformations.
However, different tendency was observed in subjective classification, MOS and ESS depending on source and target emotional states.

Compared to modern state-of-the-art emotional voice conversion methods, the performance shown in the experiment may be limited. However, StarGAN-VC promises further extension and integration considering its non-parallel many-to-many conversion characteristics.   
We believe that this promising non-parallel many-to-many voice conversion method can be used for EVC with the simultaneous use of other EVC methods to process $F_0$ and aperiodicity.
Integrating StarGAN-VC with more sophisticated EVC methods for $F_0$ and aperiodicity and applying the proposed method to sentence-level EVC will be pursued in the future.

\section{Acknowledgements}
This study is partially supported by the Japan Society for the Promotion of Science (JSPS) KAKENHI Grant-in-Aid for Scientific Research (B), grant number 18H03308, and Grant-in-Aid for Scientific Research on Innovative Areas, grant number 16H06569.

\appendix

\bibliographystyle{IEEEtran}

\bibliography{mybib}

\begin{thebibliography}{10}
\providecommand{\url}[1]{#1}
\csname url@samestyle\endcsname
\providecommand{\newblock}{\relax}
\providecommand{\bibinfo}[2]{#2}
\providecommand{\BIBentrySTDinterwordspacing}{\spaceskip=0pt\relax}
\providecommand{\BIBentryALTinterwordstretchfactor}{4}
\providecommand{\BIBentryALTinterwordspacing}{\spaceskip=\fontdimen2\font plus
\BIBentryALTinterwordstretchfactor\fontdimen3\font minus
  \fontdimen4\font\relax}
\providecommand{\BIBforeignlanguage}[2]{{%
\expandafter\ifx\csname l@#1\endcsname\relax
\typeout{** WARNING: IEEEtran.bst: No hyphenation pattern has been}%
\typeout{** loaded for the language `#1'. Using the pattern for}%
\typeout{** the default language instead.}%
\else
\language=\csname l@#1\endcsname
\fi
#2}}
\providecommand{\BIBdecl}{\relax}
\BIBdecl

\bibitem{mori2006emotional}
S.~Mori, T.~Moriyama, and S.~Ozawa, ``Emotional speech synthesis using subspace
  constraints in prosody,'' in \emph{IEEE International Conference on
  Multimedia and Expo}, 2006, pp. 1093--1096.

\bibitem{tao2006prosody}
J.~Tao, Y.~Kang, and A.~Li, ``Prosody conversion from neutral speech to
  emotional speech,'' \emph{IEEE transactions on Audio, Speech, and Language
  processing}, vol.~14, no.~4, pp. 1145--1154, 2006.

\bibitem{inanoglu2007system}
Z.~Inanoglu and S.~Young, ``A system for transforming the emotion in speech:
  Combining data-driven conversion techniques for prosody and voice quality,''
  in \emph{Eighth Annual Conference of the International Speech Communication
  Association}, 2007.

\bibitem{aihara2012gmm}
R.~Aihara, R.~Takashima, T.~Takiguchi, and Y.~Ariki, ``{GMM}-based emotional
  voice conversion using spectrum and prosody features,'' \emph{American
  Journal of Signal Processing}, vol.~2, no.~5, pp. 134--138, 2012.

\bibitem{ming2015fundamental}
H.~Ming, D.~Huang, M.~Dong, H.~Li, L.~Xie, and S.~Zhang, ``Fundamental
  frequency modeling using wavelets for emotional voice conversion,'' in
  \emph{International Conference on Affective Computing and Intelligent
  Interaction ({ACII})}, 2015, pp. 804--809.

\bibitem{Ming16-DBL}
H.~Ming, D.~Huang, L.~Xie, J.~Wu, M.~Dong, and H.~Li, ``Deep bidirectional
  {LSTM} modeling of timbre and prosody for emotional voice conversion,''
  \emph{{INTERSPEECH}}, pp. 2453--2457, 2016.

\bibitem{Luo19-EVC}
L.~Zhaojie, C.~Jinhui, T.~Tetsuya, and A.~Yasuo, ``Emotional voice conversion
  using dual supervised adversarial networks with continuous wavelet transform
  f0 features,'' \emph{IEEE/ACM Transactions on Audio, Speech, and Language
  Processing}, vol.~27, no.~10, pp. 1535--1548, 2019.

\bibitem{Choi2018stargan}
Y.~Choi, M.~Choi, M.~Kim, J.-W. Ha, S.~Kim, and J.~Choo, ``{StarGAN}: Unified
  generative adversarial networks for multi-domain image-to-image
  translation,'' in \emph{Proceedings of the IEEE conference on computer vision
  and pattern recognition}, 2018, pp. 8789--8797.

\bibitem{Kameoka18-SNM}
H.~Kameoka, T.~Kaneko, K.~Tanaka, and N.~Hojo, ``{StarGAN-VC}: Non-parallel
  many-to-many voice conversion using star generative adversarial networks,''
  in \emph{2018 IEEE Spoken Language Technology Workshop (SLT)}, 2018, pp.
  266--273.

\bibitem{Kaneko2018cyclegan}
T.~Kaneko and H.~Kameoka, ``{CycleGAN-VC}: Non-parallel voice conversion using
  cycle-consistent adversarial networks,'' in \emph{2018 26th European Signal
  Processing Conference (EUSIPCO)}.\hskip 1em plus 0.5em minus 0.4em\relax
  IEEE, 2018, pp. 2100--2104.

\bibitem{Liu2007}
K.~{Liu}, J.~{Zhang}, and Y.~{Yan}, ``High quality voice conversion through
  phoneme-based linear mapping functions with straight for mandarin,'' in
  \emph{Fourth International Conference on Fuzzy Systems and Knowledge
  Discovery (FSKD 2007)}, vol.~4, Aug 2007, pp. 410--414.

\bibitem{gao2019nonparallel}
J.~Gao, D.~Chakraborty, H.~Tembine, and O.~Olaleye, ``Nonparallel emotional
  speech conversion,'' in \emph{Proceedings of the Annual Conference of the
  International Speech Communication Association, INTERSPEECH}, 2019, pp.
  2858--2862.

\bibitem{Moriyama09}
T.~Moriyama, S.~Mori, and S.~Ozawa, ``A synthesis method of emotional speech
  using subspace constraints in prosody,'' \emph{Journal of Information
  Processing}, vol.~50, no.~3, pp. 1181--1191, 2009.

\bibitem{Morise16-WAV}
M.~Masanori, Y.~Fumiya, and O.~Kenji, ``World: a vocoder-based high-quality
  speech synthesis system for real-time applications,'' \emph{IEICE
  TRANSACTIONS on Information and Systems}, vol.~99, no.~7, pp. 1877--1884,
  2016.

\bibitem{Morise16-DAB}
M.~Masanori, ``D4c, a band-aperiodicity estimator for high-quality speech
  synthesis,'' \emph{Speech Communication}, vol.~84, pp. 57--65, 2016.

\bibitem{Russell80-ACM}
J.~A. Russell, ``A circumplex model of affect,'' \emph{Journal of personality
  and social psychology}, vol.~39, no.~6, p. 1161, 1980.

\bibitem{Kingma2014adam}
D.~P. Kingma and J.~Ba, ``Adam: A method for stochastic optimization,'' in
  \emph{International Conference on Learning Representations (ICLR)}, 2015.

\bibitem{Scherer1986-VAE}
K.~R. Scherer, ``Vocal affect expression: A review and a model for future
  research,'' \emph{Psychological bulletin}, vol.~99, no.~2, p. 143, 1986.

\end{thebibliography}

\end{document}